\baselineskip=12pt
\overfullrule=0pt

\magnification 1200
\null
\footline={\ifnum\pageno>0 \hfil \folio\hfil \else\hfill \fi}
\pageno=0

\nopagenumbers
\centerline{ }
\noindent
\centerline{\bf Quantum particle constrained to a curved surface}
\centerline{\bf in the presence of a vector potential.}

\vskip 6pt

\centerline{ Mario Encinosa and Ray H. O'Neal}
\vskip 2pt
\centerline{Department of Physics}
\vskip 1pt
\centerline{Florida A$\&$M University, Tallahassee,
Florida} \vskip 6pt

\centerline{\bf Abstract}
\vskip 3pt                        

The Schrodinger equation for a charged particle  constrained to
 a curved surface in the presence 
of a vector potential  is derived using the method of forms. 
In the limit that the particle
is brought infinitesimally close to the surface, a term arises that 
couples the component of the  vector potential normal to the surface
to the mean curvature of the surface. 

\vskip 6pt   

     It is often stated that the quasifree electrons of a nanostructure are confined 
to very nearly two dimensional regions [1]. A heterojunction well, for one example,
 confines electrons to a layer much smaller than the lateral dimensions of the
device [2], so that the physics of the object may be discussed in terms of standard
two dimensional quantum mechanical models [3]. Here we are interested 
if the reduction to  lower dimensionality 
 (even if only approximately realized) via constraints can result in novel 
effects when a charged particle
is subject to a static vector potential. There are none
for a flat surface from elementary considerations;
 however, it has been shown that  when a particle in three dimensional space
is constrained  to a two dimensional  curved surface,  curvature terms enter 
the Schrodinger equation (or the path integral) for the particle [4,5,6,7,8].
In this letter, the methods of [9] are extended to include a vector potential.

Let $d$ be the standard exterior derivative operator that provides a mapping from
 $p$ forms to $p+1$ forms appropriate to a coordinate
system $(q_1,q_2,q)$. The minimal substitution rule for the covariant
 derivative of a negatively charged
particle in the presence of a vector potential [10] gives
$$
{1 \over i} D \rightarrow {1 \over i} \big [ d + ieA \big ]
\eqno (1) 
$$
In the absence of a scalar potential, A is 
$$
A = A_1 \sigma_1 + A_2 \sigma_2 + A_3 \sigma_3
\eqno(2)
$$
\noindent
with  $\sigma_i$  one forms derived from applying $d$ to the Monge form 
${\bf x}(q_1,q_2,q)$ for a given coordinate system.
For $\Psi$ a zero form function, the Schrodinger equation in this notation
(in natural units) becomes
$$
-{1 \over 2}* [D *D \Psi] = i \hbar {\partial \Psi \over \partial t}
\eqno(3)
$$
\noindent
with $*$ the Hodge star operator[11]. In a gauge with $d * A  = 0$, eq.(3) may
be written in the familiar form
$$
-{1 \over 2} *\bigg [d *d + 2ieA*d  - (A*A) \bigg]\Psi = i \hbar {\partial \Psi \over \partial t}.
\eqno(4)
$$
\noindent
Because the main interest here is in regions near to and on surfaces, ${\bf x}$ 
is set equal to
a surface part with coordinates $(q_1,q_2)$ plus a part everywhere normal to
the surface with coordinate $q$,
$$
{\bf x}(q_1,q_2,q) = {\bf r}(q_1,q_2) + q {\bf e}_3
\eqno(5)
$$
so that
$$
d {\bf x} = d{\bf r} + dq {\bf e}_3 + q d{\bf e}_3
$$
$$
\qquad = \sigma_1 {\bf e}_1 + \sigma_2 {\bf e}_2 + \sigma_3 {\bf e}_3.
\eqno(6)
$$
To avoid unneccessary complexity, consider an axially symmetric 
surface given by the set of points that satisfy the surface part
 of eq. (5),
$$
{{\bf r}(\rho,\phi)} = \rho \ {\rm cos}\phi \ {\bf i} + \rho \  {\rm sin}\phi \ {\bf j}
 + S(\rho) \ {\bf k}.
\eqno (7)
$$
\noindent
In terms of a local basis, points near
the surface  can be written either in terms of eq.(5) or in terms of
a local set of unit vectors as (subscripts on $S$ denote differentiation)
$$
{{\bf x}(\rho,\phi,q)} = {{\bf r}(\rho,\phi)}+q{\bf e}_3 = 
{1 \over Z} {\big [} (\rho + S S_{\rho}){\bf e}_1
+ (S - \rho S_{\rho}+ q){\bf e}_3 {\big ]}
\eqno (8)
$$
\noindent
with
$$
Z = {\sqrt {1 + {S_{\rho}}^2}},
\eqno (9)
$$
$$
{\bf e}_1 = {1 \over Z} {\big [} {\rm cos} \phi \ {\bf i} +  {\rm  sin}\phi \ {\bf j}
 + S_{\rho} \  {\bf k} {\big ]},
\eqno (10)
$$
$$
{\bf e}_2 = -{\rm sin}\phi \ {\bf i} + {\rm cos}\phi \ {\bf j} ,
\eqno (11)
$$
and
$$
{\bf e}_3 = {1 \over Z} {\big [} -S_{\rho} {\rm cos}\phi \ {\bf i}  
- S_{\rho} {\rm sin}\phi \ {\bf j} + {\bf k} {\big ]}.
\eqno (12)
$$
\noindent
The one forms for this system can be read off from
$$
d{\bf x} = d\rho \ Z {\big [}1 -{q S_{\rho\rho} \over Z^3}{\big ]}{\bf e}_1 
+ d\phi\rho {\big [}1 -{q S_{\rho} \over Z \rho}{\big ]} {\bf e}_2
 + dq \ {\bf e}_3 \equiv \sigma_1{\bf e}_1  + \sigma_2{\bf e}_2 
 + \sigma_3{\bf e}_3.
\eqno(13)
$$
\noindent
It is possible to write  eq.(4) in this geometry, but
there is little to be gained by writing the general expression 
in full detail since
 our primary interest here is the modification of eq.(4) when the particle
is constrained to the surface.
A constraint that brings the particle to the surface
can be thought of as being effected by  a large confining potential
everywhere normal to the surface, i.e., a local function of $q$ in
the ${\bf e}_3$ direction. The usual choice [5,6] for this term is
$$
V_n(q) = {1 \over 2} \omega^2 q^2
\eqno(14)
$$
with eventually $\omega \rightarrow \infty$ and $q \rightarrow 0$.
As the particle approaches the surface,
 we anticipate a decoupling of the wavefunction into tangential
 and normal degrees of freedom 
$$
{\Psi} (\rho , \phi , q) \  {\rightarrow} \  \chi_t (\rho , \phi) \chi_n (q).
$$
A consistent relation for the norm is obtained by insisting on the condition [4]
$$
\vert \Psi {\vert} ^2  F  dS dq  =  \vert {\chi}_t {\vert}^2 dS
                \vert {\chi}_n {\vert}^2 dq, 
\eqno(15)
$$
\noindent
so that
$$
\Psi = {{\chi_t \chi_n} \over {\sqrt F}}
\eqno(16)
$$
\noindent
with 
$$
 F = 1 + 2 q_3 H + q_3^2 K 
\eqno(17)
$$
\noindent
and H, K the mean and Gaussian curvatures of the surface. Their explicit forms
are given by
$$
H = -{1 \over 2}\bigg[ { S_{\rho} \over {Z\rho}} + {S_{\rho\rho} \over {Z^3}} \bigg],
\eqno(18)
$$
and
$$
K =  {S_{\rho} S_{\rho\rho}  \over {\rho Z^4}}.
\eqno(19)
$$
\noindent
The ansatz of eq.(16) can be inserted into eq.(4) to give the following
relations for the tangential and normal functions in the $q \rightarrow 0$ limit:
$$
-{1 \over 2}  \bigg [ Z^2 
\big ( {\partial^2 \chi_t \over \partial \rho^2} 
+ {1 \over \rho}{\partial \chi_t\over \partial \rho} \big )\chi_n  +
 {1 \over \rho^2}{\partial^2 \chi_t \over \partial \phi^2}\chi_n 
+ (H^2-K)\chi_t\chi_n  + {\partial^2 \chi_n \over \partial  q^2}\chi_t
 - Z^4 S_{\rho} S_{\rho \rho} {\partial \chi_t \over \partial \rho}\chi_n \bigg ]
$$
$$
-  ie \bigg [ {A_1 \over Z} {\partial \chi_t\over \partial \rho}\chi_n
+ {A_2 \over \rho} {\partial \chi_t \over \partial \phi}\chi_n
 + A_3{\partial \chi_n \over \partial q}\chi_t - A_3H \chi_t\chi_n\bigg ]
$$
$$
+{e^2 \over 2} \bigg [ {A_1}^2 + {A_2}^2 + {A_3}^2 \bigg ]\chi_t\chi_n
+ V_n(q) \chi_t\chi_n = 
i\hbar {\partial \over \partial t} \chi_t\chi_n
\eqno(20)
$$
\noindent
The $(H^2-K)$ term appearing in the first grouping has been discussed
elsewhere [9,12] and will not prove important for what follows. 
It is interesting to note that a coupling 
of $A_3$ to the mean curvature $H$ obtains even if a limiting procedure 
does not occur;
it originates from a term proportional to  ${\partial F \over \partial q}$
and is present only if  there is  variation of the unit normal ${\bf e}_3$.
 Should the $q \rightarrow 0$ condition not be imposed,
higher order curvature terms would be present in eq.(20).

As an admittedly unphysical example of how the curvature term can produce 
a nontrivial consequence, consider a  situation with $A_1 = A_2 = 0$, and
$A_3, H$ both functions of $\rho$ only. Next for 
$$
V_n(q) >> A_3 {\partial  \over \partial q} ln [ \chi_n(q)]
\eqno(21)
$$
\noindent
a separable pair of equations obtain,
$$
-{1 \over 2}  \bigg [ Z^2 
\big ( {\partial^2 \chi_t \over \partial \rho^2} 
+ {1 \over \rho}{\partial \chi_t\over \partial \rho} \big )  -
 {m^2 \over \rho^2}\chi_t 
- Z^4 S_{\rho} S_{\rho \rho} {\partial \chi_t \over \partial \rho} + (H^2-K)\chi_t
 \bigg ]
$$
$$
 + ie A_3 H \chi_t +{e^2 \over 2} {A_3}^2\chi_t
= E_t \chi_t
\eqno(22)
$$
\noindent
and
$$
-{1 \over 2}{\partial^2 \chi_n \over \partial  q^2} + V_n(q) \chi_n = E_q \chi_n
\eqno(23)
$$
\noindent
with $m$ an aziumthal quantum number. Suppose that the product of 
$A_3$ and $H$  over some region $\Gamma$ is approximately constant and zero
elsewhere. Then for whatever $\chi_t$
 results from the solving the tangential equation, the solution in $\Gamma$
would become
$$
\chi_t \rightarrow \chi_t \ {\rm exp}[\pm e |A_3 Ht|].
\eqno(24)
$$
\noindent
 Eq. (24) illustrates a peculiar situation wherein the sign of
a term depending on curvature can have either dissapative or pathological
behavior. Cleary this situation is not expected to persist for physically realizable
systems, but it serves to show that peculiar results can obtain when studying
constrained quantum systems.

In this letter, differential forms were used to derive the Schrodinger equation
for a three dimensional particle constrained to a two dimensional cylindrically 
symmetric surface in the presence of a static vector potential.
For arbitrary choices of geometry and field configuration, it is  difficult 
to find closed form solutions of eq.(20). Nevertheless, the above discussion
indicates that the interplay between surface geometry and applied fields may
be important.
It is worth reemphasizing that even for finite $q$ there is nontrivial 
coupling of $A$ to surface terms through differentiations of eq.(17). 
This is a manifestation of a general result;
  imposing constraints in three dimensional space to restrict particle
dynamics to a two dimensional space   gives
different results than an ${\it a \ priori}$ two dimensional model.

\vfill \eject
\vskip 6pt
\centerline{\bf Acknowledgments}
\vskip 2pt
The author would like to acknowledge useful discussions with
 Ray O'Neal and Lonnie Mott.

\vfill\eject
\vskip 6pt
\centerline{\bf REFERENCES}
\vskip 6pt
\noindent
1. R. Ashoori, ${\it Nature}$, vol. 378, pp. 413-419, 1998.
\vskip 6pt
\noindent
2. J. R. Chelilowsky and S. G. Louie, eds. ${\it Quantum\ Theory\ of\ Real\ Materials}$,
(Kluwer Academic Publishers, Massachusetts, 1996).
\vskip 6pt
\noindent
3. D. Pfannkuche, R. Gerhardts, P. Maksym andV. Gudmundsson, 
Phys. B  ${\bf 189}$, 6, (1993).
\vskip 6pt
\noindent
4.  R. C. T. da Costa, Phys. Rev. A ${\bf 23}$, 1982 (1981).
\vskip 6pt
\noindent
5.  R. C. T. da Costa, Phys. Rev. A ${\bf 25}$, 2893 (1982).  
\vskip 6pt
\noindent
6.  M. Burgess and B. Jensen, Phys. Rev. A ${\bf 48}$, 1861 (1993).
\vskip 6pt
\noindent
7.  S. Matsutani, J. Phys. Soc. Japan 61, ${\bf 55}$, (1992).
\vskip 6pt
\noindent
8.  S. Matsutani and H. Tsuru, J. Phys. Soc. Japan, ${\bf 60}$,
 3640 (1991).
\vskip 6pt
\noindent
9. M. Encinosa and B. Etemadi, Physical Review A ${\bf 58}$, 77 (1998).
\vskip 6pt
\noindent
10. B. Skarsgard, ${\it Geometry,\ Particles \ and \ Gauge \ Fields}$  
(Springer Verlag, New York, 1998).
\vskip 6pt
\noindent
11.   H. Flanders; ${\it Differential\ Forms \ with \ Applications \ to \ the 
\ Physical \ Sciences}$  
(Dover Books, New York, 1989).
\vskip 6pt
\noindent
12. M. Encinosa and B. Etemadi, Physica  B, {\bf 266 }, 361 (1999). 
\vskip 6pt
\noindent

\end